\documentclass[aps,prd,superscriptaddress,preprint]{revtex4-1}
\usepackage{amsmath,amssymb,bm}
\usepackage{graphicx,graphics,color}
\usepackage[colorlinks=true,citecolor=blue,linkcolor=blue,urlcolor=blue]{hyperref}
\usepackage{epsfig}
\usepackage{latexsym}
\usepackage{physics}
\usepackage{rotating}
\usepackage{subfigure}
\usepackage{centernot}
\usepackage{longtable}
\usepackage{hhline,multirow,tabularx}
\usepackage{float}
\usepackage{mathrsfs}
\usepackage{xr}
\usepackage{slashed}
\usepackage{adjustbox}
\usepackage{soul}

\newcommand{\RNum}[1]{\uppercase\expandafter{\romannumeral #1\relax}}

\newcommand{\yb}{\bar{y}}

\begin{document}
\preprint{ADP-21-18/T1165, JLAB-THY-22-3576}

\title{Helicity-dependent distribution of strange quarks in the proton \\ from nonlocal chiral effective theory}

\author{Fangcheng He}
\affiliation{CAS Key Laboratory of Theoretical Physics, Institute of Theoretical Physics, CAS, Beijing 100190, China}
\author{Chueng-Ryong Ji}
\affiliation{\mbox{Department of Physics, North Carolina State University, Raleigh, North Carolina 27695, USA}}
\author{W. Melnitchouk}
\affiliation{Jefferson Lab, Newport News, Virginia 23606, USA}
\author{Y.~Salamu}
\affiliation{\mbox{School of Physics and Electrical Engineering, Kashi University, Kashi, 844000, Xinjiang, China}}
\author{A. W. Thomas}
\affiliation{CSSM and ARC Centre for Dark Matter Particle Physics, Department of Physics, University of Adelaide, Adelaide SA 5005, Australia}
\author{P. Wang}
\affiliation{Institute of High Energy Physics, CAS, Beijing 100049, China}
\affiliation{\mbox{College of Physics Sciences, University of Chinese Academy of Sciences, Beijing 100049, China}}
\author{X. G. Wang}
\affiliation{CSSM and ARC Centre for Dark Matter Particle Physics, Department of Physics, University of Adelaide, Adelaide SA 5005, Australia}

\begin{abstract}
The helicity-dependent strange quark distribution in the proton, $\Delta s$, is calculated in a nonlocal chiral SU(3) effective field theory.
The hadronic proton to meson plus octet or decuplet baryon splitting functions are derived at the one-loop level, with loop integrals rendered finite by correlation functions introduced in the nonlocal Lagrangian.
Within the convolution framework, the proton strange helicity distribution is obtained using spin-flavor symmetry to constrain the input valence quark distributions in the hadronic intermediate states. 
The polarized strange quark distribution is found to be quite small, with the lowest moment of $\Delta s$ negative, but consistent with recent global QCD analyses.
\end{abstract}

\date{\today}
\maketitle

\section{Introduction}
\label{sec:intro}

It has been over 30 years since the European Muon Collaboration (EMC) published their polarized deep-inelastic scattering (DIS) measurement of the proton's spin-dependent structure function, $g_1$~\cite{Ashman:1987hv}.
The result suggested that only a very small fraction of the proton's spin was carried by quarks --- an initially shocking discovery which contradicted the prevailing quark model view in which constituent quarks accounted for the proton's global quantum numbers, including its spin. 
Subsequent experiments with increasing precision and kinematic reach were performed at SLAC~\cite{Anthony:1996mw, Abe:1998wq, Abe:1997cx, Anthony:2000fn, Anthony:1999py, Anthony:2002hy}, HERMES~\cite{Ackerstaff:1997ws, Airapetian:2006vy, Airapetian:2011wu}, SMC~\cite{Adeva:1998vv, Adeva:1999pa}, COMPASS~\cite{Alexakhin:2006oza, Alekseev:2010hc}, Jefferson Lab~\cite{Zheng:2003un, Dharmawardane:2006zd, Prok:2008ev, Prok:2014ltt, Guler:2015hsw, Fersch:2017qrq, Parno:2014xzb, Posik:2014usi, Solvignon:2013yun, Armstrong:2018xgk}, and RHIC~\cite{Adamczyk:2014xyw, Adare:2015gsd, Adam:2019aml}, and various  global QCD analyses of these data in terms of spin-dependent parton distribution functions (PDFs) have been carried out~\cite{deFlorian:2009vb, deFlorian:2014yva, Hirai:2008aj, Blumlein:2010rn, Leader:2011tm, Leader:2014uua, Khanpour:2017cha, Nocera:2014gqa, Sato:2016tuz, Sato:2016wqj, Ethier:2017zbq, Zhou:2022wzm}.
A recent analysis from the JAM Collaboration, for instance, gives a total fraction $\Delta\Sigma = 0.36 \pm 0.09$ of the proton's spin carried by quarks at a scale of $Q^2=1$~GeV$^2$~\cite{Ethier:2017zbq}.

One early explanation proposed for the small value of $\Delta\Sigma$ was that the contribution to the proton spin from strange quarks, which is much less well determined than that from up and down quarks, was large and negative.
Within the assumptions traditionally made in phenomenological analyses, such as SU(3) flavor symmetry and the equivalence of the strange and antistrange polarizations, $\Delta s = \Delta \bar{s}$, the integrated strange quark polarization has typically come in at around $\Delta s^+ \equiv \Delta s + \Delta \bar s \approx -0.1$.
In this scenario, the nonsinglet axial charge, $a_8$, is extracted from hyperon beta decays to be 
    $a_8 = \Delta u^+ + \Delta d^+ - 2 \Delta s^+
         = 0.58 \pm 0.03$ \cite{Close:1993mv},
in which case a strange quark polarization of $\approx -0.1$ would give a total quark spin contribution
    $\Delta\Sigma \equiv \Delta u^+ + \Delta d^+ + \Delta s^+
        = a_8 + 3 \Delta s^+$
that would be close to the phenomenological result.

The accuracy of the flavor SU(3) symmetry assumption has been questioned, on the other hand, in several analyses~\cite{Jaffe:1989jz, Ratcliffe:2004jt, Bass:2009ed} that have suggested that the uncertainty could be as large as $\approx 20\%$.
A re-evaluation of the nucleon's axial-charges in the cloudy bag model~\cite{Thomas:1981vc, Thomas:1982kv}, for example, taking into account the effect of the one gluon exchange hyperfine interaction and the meson cloud, led to the value $a_8 = 0.46 \pm 0.05$~\cite{Bass:2009ed}.
Recent lattice simulations directly including the effects of disconnected quark loops have yielded smaller magnitudes for the strange quark polarization,
    $\Delta s^+_{\rm latt} = -0.046 \pm 0.008$~\cite{Alexandrou:2020sml},
while an analysis of the proton spin taking into account the angular momentum carried by the pion cloud~\cite{Schreiber:1988uw, Thomas:2008ga, Myhrer:2007cf} favors a value $\approx -0.01$~\cite{Bass:2009ed, Yamaguchi:1989sx}.
The recent JAM global QCD analysis~\cite{Ethier:2017zbq}, which used data from inclusive and semi-inclusive DIS in order to relax the SU(3) symmetry constraint, also supports a smaller magnitude for the strange quark polarization,
    $\Delta s^+_{\rm JAM} = -0.03 \pm 0.10$,
at a scale $Q^2=1$~GeV$^2$, but with a somewhat larger uncertainty.
For a review of the status of global QCD analyses and lattice QCD simulations, see Ref.~\cite{Lin:2017snn}, while for an explanation of the importance of the mismatch between the scale appropriate to quark models and that of lattice QCD and DIS, see Ref.~\cite{Thomas:2008ga}.

On the theoretical front, considerable progress has been made in the last few years in developing the formalism and feasibility of extracting the momentum dependence of quark distributions from lattice QCD calculations of quasi-PDFs and pseudo-PDFs~\cite{Alexandrou:2019lfo, Joo:2019jct}, including contributions from the $q\bar q$ sea, and exploring ways in which lattice data could constrain the phenomenological distributions~\cite{Bringewatt:2020ixn}.
For continuum-based approaches, early model-dependent work focused on the effects of proton fluctuations to kaon-hyperon intermediate states on strange nucleon observables~\cite{Signal:1987gz, Melnitchouk:1996fj, Melnitchouk:1999mv, Holtmann:1996be, Zamani:2001tn, Cao:2003zm}, although reliably quantifying such effects has proven challenging.
A more systematic methodology for quantifying the effects of virtual meson loops on PDFs~\cite{Thomas:1983fh} was formulated subsequently in the framework of chiral effective field theory (EFT), and used to study the unpolarized light quark asymmetry $\bar d - \bar u$, the strange--antistrange asymmetry $s - \bar s$, as well as the strange quark helicity in the \mbox{proton~\cite{Wang:2020hkn, Salamu:2014pka, Wang:2016eoq, Wang:2016ndh}}.

Along these lines, a nonlocal chiral effective theory was recently proposed, which allows the study of hadron properties at relatively large momentum transfer~\cite{Wang:2014tna, He:2017viu, He:2018eyz}, while consistently taking into account the finite size of hadrons from the underlying chiral Lagrangian.
This framework was used to compute electromagnetic 
form factors of the nucleon~\cite{He:2017viu, He:2018eyz}, as well as collinear parton distributions~\cite{Salamu:2018cny, Salamu:2019dok} and transverse momentum dependent distributions~\cite{He:2019fzn}, such as the Sivers function, for the sea quarks in the nucleon.
Furthermore, if the nonlocal behavior is assumed to be a general property of all the interactions, it produces interesting results when applied to the lepton anomalous magnetic moments~\cite{He:2019uvu}.

In this paper, we extend our previous analysis~\cite{Salamu:2019dok, He:2019fzn} of the chiral loop contributions to the nonperturbative strange quark PDF in the polarized sector within the framework of nonlocal chiral effective theory.
In Sec.~\ref{sec:lagrangian}, we review the derivation of the nonlocal chiral Lagrangian starting from the local effective Lagrangian.
From this the spin-dependent hadronic splitting functions are computed in Sec.~\ref{sec:fy}, for the case of a covariant dipole form factor, and numerical results are presented in Sec.~\ref{sec:results}.
Finally, Sec.~\ref{sec:conclusion} contains a summary of our findings and suggestions for future research.

\section{Effective Lagrangian}
\label{sec:lagrangian}

Our analysis is based on the chiral SU(3)$_L\times$SU(3)$_R$ effective Lagrangian, describing interactions of octet ($B$) and decuplet ($T_\mu$) baryons with pseudoscalar mesons ($\phi$) 
\cite{Jenkins:1990jv, Bernard:2007zu, Hacker:2005fh},
\begin{eqnarray}
\label{eq:ch8}
{\cal L}
&=&   {\rm Tr} \big[ \bar B (i\slashed{D} - M_B) B \big]
 - \frac{D}{2}\, {\rm Tr} \big[ \bar B \gamma^\mu \gamma_5 \{u_\mu, B\} \big]
 - \frac{F}{2}\, {\rm Tr} \big[ \bar B \gamma^\mu \gamma_5 [ u_\mu ,B ] \big]
\nonumber\\
&-& \frac{F-D}{2}\, {\rm Tr} \big[ \bar B \gamma^\mu \gamma_5 B \big]
    {\rm Tr} \big[ u_\mu\big ]
 - \frac{{\cal C}}{2}
   \left( \epsilon^{ijk}\, \overline{T}_\mu^{ilm}
	  \Theta^{\mu\nu} (u_\nu)^{lj} B^{mk} + {\rm H.c.}
   \right)
\nonumber\\
&+&\overline{T}_\mu^{ijk}
   \big( i\gamma^{\mu \nu \alpha} D_\alpha - M_T \gamma^{\mu\nu} \big) T_\nu^{ijk}
- \frac{{\cal H}}{2}\,
    \overline{T}_\mu^{ijk} \gamma^{\mu\nu\alpha} \gamma_5 (u_\alpha)^{kl}\, T_\nu^{ijl},
\end{eqnarray}
where $M_B$ and $M_T$ are the masses of the octet and decuplet baryons, and $D$, $F$, $\cal C$ and $\cal H$ denote the baryon-meson coupling constants.
%
%
The octet-decuplet baryon transition operator $\Theta^{\mu\nu}$ is given by
\begin{equation}
\Theta^{\mu\nu}
= g^{\mu\nu} - \big( Z + \tfrac12 \big) \gamma^\mu \gamma^\nu
\label{eq:Theta},
\end{equation}
where $Z$ is the decuplet off-shell parameter (usually chosen to be $Z=-1/2)$.
The pseudoscalar mesons couple to baryons through the vector and axial vector combinations involving the field 
    $u = \exp \big( i \phi / \sqrt2\, f \big)$, 
\begin{eqnarray}
\Gamma_\mu
&=& \frac12
    \left( u^\dagger \partial_\mu u + u \partial_\mu u^\dagger
    \right)
 +  \frac{i}{2}
    \left( u^\dagger \lambda^a u - u\lambda^a u^\dagger
    \right) a_\mu^a,
\\
u_\mu
&=& i\left( u^\dagger \partial_\mu u - u \partial_\mu u^\dagger \right)
 -  \left( u^\dagger \lambda^a u + u\lambda^a u^\dagger
    \right) a_\mu^a,
\label{eq:22}
\end{eqnarray}
where $f$ is the pseudoscalar decay constant, $a_\mu^a$ corresponds to the external axial-vector fields, and $\lambda^a$ ($a=0, \ldots, 8$) represent the unit matrix and the eight Gell-Mann matrices.
For further details about the SU(3) chiral Lagrangian see the discussions in Ref.~\cite{Jenkins:1990jv, Bernard:2007zu, Hacker:2005fh, Salamu:2018cny}.

From the chiral Lagrangian (\ref{eq:ch8}) one can derive the electromagnetic currents that couple to the external field $a_\mu^a$,
\begin{eqnarray}
J_A^{\mu,a}
&=& \frac12 {\rm Tr}
   \big[
   \bar B \gamma^\mu
   [ u \lambda^a u^\dagger - u^\dagger \lambda^a u, B ]	
   \big]
      + \frac{D}{2}\, {\rm Tr}
   \big[
   \bar B \gamma^\mu \gamma_5
   \left\{ u \lambda^a u^\dagger + u^\dagger \lambda^a u, B
   \right\}
   \big]						\notag\\
&+&
   \frac{F}{2}\, {\rm Tr}
   \big[
   \bar B \gamma^\mu \gamma_5
   [ u \lambda^a u^\dagger + u^\dagger \lambda^a u, B ]
   \big]				
+ \frac{F-D}{2}\, {\rm Tr}
   \big[ \bar B \gamma^\mu \gamma_5 B \big ]
   {\rm Tr} \big[ u \lambda^a u^\dagger + u^\dagger \lambda^a u  \big]	\notag\\
&+& \frac{\cal C}{2}\,
   \big(
   \overline{T}_\nu \Theta^{\nu\mu}
   (u \lambda^a u^\dagger + u^\dagger \lambda^a u) B
   + {\rm H.c.}
   \big)						
+  \frac{\cal H}{2}\,
   \overline{T}_\nu \gamma^{\nu\alpha\mu}
   \left( u \lambda^a u^\dagger + u^\dagger \lambda^a u, T_\alpha
   \right),
\label{eq:ch1}
\end{eqnarray}
where the notations are as defined in Ref.~\cite{Salamu:2018cny}.
%
%
Following the methodology discussed in Refs.~\cite{Terning:1991yt, Holdom:1992fn, Abe:1997cx, Faessler:2003yf, Wang:1996zu, He:2017viu, He:2018eyz}, we write the nonlocal baryon--meson interaction Lagrangian for the meson coupling to a proton as~\cite{Salamu:2018cny}
\begin{eqnarray}
{\cal L}^{\rm (nonloc)}_{\rm had}(x)
&=& \bar{p}(x)
  \bigg(
    \frac{C_{B\phi}}{f} \gamma^\mu \gamma^5 B(x)\,
  + \frac{C_{T\phi}}{f} \Theta^{\mu\nu} T_\nu(x)
  \bigg)
  \int \dd^4{a} F(a)\, \partial_{\mu}
  \phi(x+a) + {\rm H.c.}		
\notag\\
&+&
  \frac{i C_{\phi\phi^\dag}}{2 f^2}\,
  \bar{p}(x) \gamma^\mu p(x)                                  
  \int \dd^4{a}\, F(a)\, \phi(x+a) 
  \int \dd^4{b}\, F(b)\, \big( \partial_\mu \phi^\dag(x+b) + {\rm H.c.} \big),~~
\label{eq:j4}
\end{eqnarray}
where $C_{B\phi}$, $C_{T\phi}$ and $C_{\phi\phi^\dagger}$ are the coupling constants for the $pB\phi$, $pT\phi$ and $pp\phi\phi^\dagger$ interactions, respectively (see Table~I of Ref.~\cite{Salamu:2018cny}).

The nonlocal interaction between a quark $q$ in a hadron and the external axial-vector field $a_\mu$ is given by
\begin{eqnarray}
{\cal L}^{\rm (nonloc)}_{q\, \rm ext}(x)
&=& \int \dd^4{a}\, F(a)
  \Big(
    C_B^q\, \bar B(x) \gamma^\mu \gamma^5 B(x)\,
  + C_T^q\, \overline{T}_\alpha \gamma^\mu\gamma^5 T^\alpha(x)
  \Big)\,
  a_\mu(x+a) 
\notag\\
&+& C_{BT}^q\, \int \dd^4{a}\, F(a)
  \Big( \bar B(x)\, \Theta^{\mu\nu}T_\nu(x) + {\rm H.c.} \Big)\,
  a_\mu(x+a)
\notag\\
&+& \frac{iC_{B\phi}^q}{f} 
  \int \dd^4{a}\, F(a) \int \dd^4{b}\, F(b)\,
  \Big( \bar p (x) \gamma^\mu B(x)\, \phi(x+a)+ {\rm H.c.} \Big)\,
  a_\mu(x+b)
\notag\\
&+& \frac{C_{\phi\phi^\dag}^q}{f^2}
  \int \dd^4{a}\, F(a) \int \dd^4{b}\, F(b)\int \dd^4{c}\, F(c)\,
  \bar{p}(x) \gamma^\mu \gamma^5 p(x)\,
\notag\\
& & \hspace*{5.5cm} \times\
  \phi(x+a)\, \phi^\dag(x+b)\,
  a_\mu(x+c),
\label{eq:ex}
\end{eqnarray}
where $C_j^q$ ($j=B$, $T$, $BT$, $B\phi$, or $\phi\phi^\dagger$) are the coupling constants (axial charges) for the interaction between the quark $q$ in the hadronic configuration $j$ and the axial-vector field.
In the present work we focus on the strange quark contribution, for which the  corresponding couplings $C_j^s$ are listed in Table~\ref{tab:axial-charge}.

\begin{table}[t]
\begin{center}
\caption{Coupling constants $C_B^s$, $C_T^s$, $C_{BT}^s$, $C_{B\phi}^s$ and $C_{\phi\phi^\dag}^s$ for the interaction between a strange quark $s$ and an external axial-vector field in the corresponding hadronic configurations.\\}
\begin{tabular}{c|cccccc}
\hline\hline
  \hspace*{0.5cm}\boldmath{$B$}\hspace*{0.5cm}
& \hspace*{0.5cm}\boldmath{$\Lambda$}\hspace*{0.5cm}
& \hspace*{0.5cm}\boldmath{$\Sigma^0$}\hspace*{0.5cm}
& \hspace*{0.5cm}\boldmath{$\Sigma^+$}\hspace*{0.5cm}
\\
  \hspace*{0.5cm}$C_{B}^s$\hspace*{0.5cm}
& \hspace*{0.5cm}$F+\frac13 D$\hspace*{0.5cm}
& \hspace*{0.5cm}$F-D$\hspace*{0.5cm}
& \hspace*{0.5cm}$F-D$\hspace*{0.5cm} 
\\ \hline
  \hspace*{0.5cm}\boldmath{$T$}\hspace*{0.5cm}
& \hspace*{0.5cm}\boldmath{$\Sigma^{*0}$}\hspace*{0.5cm}
& \hspace*{0.5cm}\boldmath{$\Sigma^{*+}$}\hspace*{0.5cm}
\\
  \hspace*{0.5cm}$C_{T}^s$\hspace*{0.5cm}
& \hspace*{0.5cm}$\frac13 {\cal H}$\hspace*{0.5cm}
& \hspace*{0.5cm}$\frac13 {\cal H}$\hspace*{0.5cm} 
\\ \hline
  \boldmath{$BT$}
& \boldmath{$\Sigma^0 \Sigma^{*0}$}
& \boldmath{$\Sigma^+ \Sigma^{*+}$} 
\\
$C_{BT}^s$
& $\frac13 {\cal C}$
& $-\frac13{\cal C}$
\\ \hline
  \boldmath{$B\phi$}
& \boldmath{$\Lambda K^+$}
& \boldmath{$\Sigma^0 K^+$}
& \boldmath{$\Sigma^+ K^0$} 
\\
$C_{B\phi}^s$
& $\frac{\sqrt3}{2} $
& $\frac12$ 
& $\frac{1}{\sqrt2}$ 
\\ \hline
  \boldmath{$\phi\phi^\dag$}
& \boldmath{$K^0 \overline{K}^0$}
& \boldmath{$K^+K^-$} 
\\
$C_{\phi\phi^\dag}^s$
& $\frac12 (F-D)$
& $F$ 
\\
\hline\hline
\end{tabular}
\label{tab:axial-charge}
\end{center}
\end{table}

For the electromagnetic form factors and unpolarized PDFs discussed in Refs.~\cite{He:2017viu, He:2018eyz, Salamu:2018cny} the path integral of the vector field in the gauge link was necessary to guarantee local gauge invariance.
In the current application to interactions with the axial-vector field and spin-dependent PDFs, there is no associated conserved charge and hence no gauge link term in the nonlocal Lagrangian.
The corresponding nonlocal axial-vector current can be obtained from Eq.~(\ref{eq:ex}) and is given by 
\begin{eqnarray}
J_A^{\mu,q}(x)
&=& \int \dd^4{a}\, F(a)\,
\Big(
    C_B^q\, \bar B(x-a) \gamma^\mu \gamma^5 B(x-a)\,
  + C_T^q\, \overline{T}_\alpha(x-a) \gamma^\mu\gamma^5\, T^\alpha(x-a)
\Big)
\notag\\
&+& C_{BT}^q \int \dd^4{a}\, F(a)\,
\Big( 
    \bar B(x-a)\, \Theta^{\mu\nu}\, T_\nu(x-a) + {\rm H.c.} 
\Big)
\notag\\
&+& \frac{iC_{B\phi}^q}{f} 
\int \dd^4{a}\, F(a)
\int \dd^4{b}\, F(b)
\Big(
    \bar{p}(x-b) \gamma^\mu B(x-b)\, \phi(x+a-b) + {\rm H.c.}
\Big)\, 
\notag\\
&+& \frac{C_{\phi\phi^\dag}^q}{f^2}
\int \dd^4{a}\, F(a)
\int \dd^4{b}\, F(b)
\int \dd^4{c}\, F(c)\, \bar{p}(x-c) \gamma^\mu \gamma^5\, p(x-c) 
\notag\\
& & \hspace*{5.5cm}
\times\ \phi(x+a-c)\, \phi^\dag(x+b-c).
\label{eq:Jem}
\end{eqnarray} 
From the nonlocal axial current and Lagrangian, in the next section we will compute the proton to baryons + meson splitting functions necessary for the helicity-dependent strange quark distribution.

\section{Hadronic splitting functions}
\label{sec:fy}

\begin{figure}[t]
\vspace{0.5cm}
\center
\includegraphics[width=0.9\columnwidth]{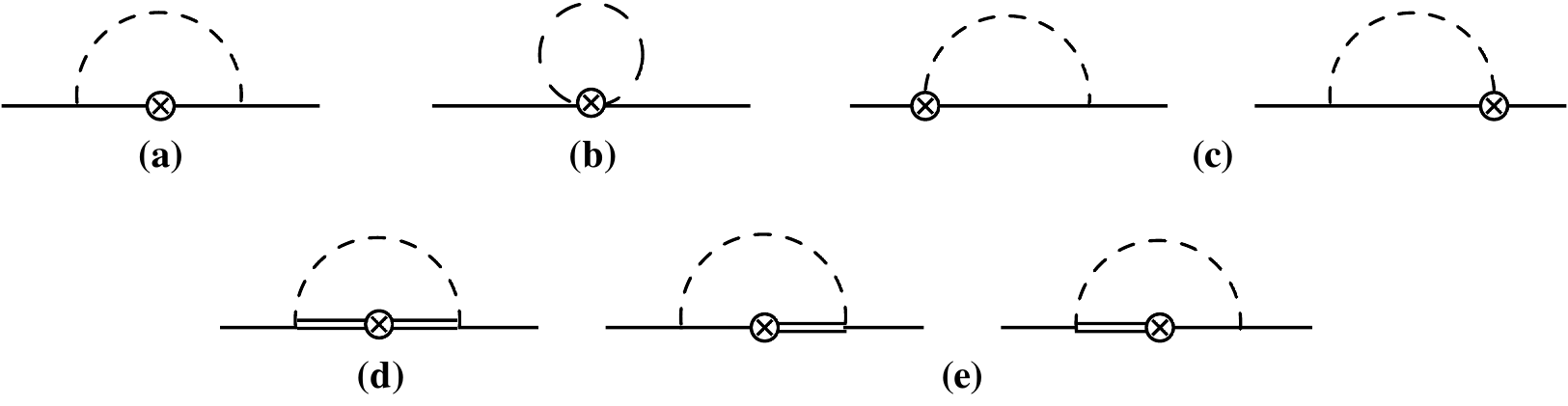}
\caption{One-loop contributions to the spin-dependent PDFs of the nucleon from 
	{\bf (a)} octet rainbow,
	{\bf (b)} tadpole,
	{\bf (c)} Kroll-Ruderman,
	{\bf (d)} decuplet rainbow, and
	{\bf (e)} octet-decuplet transition diagrams. 
The octet baryons, decuplet baryons and pseudoscalar mesons are represented by the solid, double-solid and dashed lines, respectively, while the symbol $\otimes$ denotes insertion of the hadronic axial current operator in Eq.~(\ref{eq:Jem}).}
\label{fig:loop-octet-decuplet}
\end{figure}

The spin-dependent hadronic splitting functions, $\Delta f_j$, can be evaluated from the matrix elements of the hadronic operators of the axial current, which correspond to the one meson loop diagrams in Fig.~\ref{fig:loop-octet-decuplet} \cite{Salamu:2018cny, Wang:2020hkn}.
The matrix elements of the hadronic operators give rise to the octet rainbow, tadpole, Kroll-Ruderman (KR), decuplet rainbow, and octet-decuplet transition splitting functions, as illustrated by the diagrams in Fig.~\ref{fig:loop-octet-decuplet}.
A detailed derivation of the splitting functions corresponding to the processes illustrated in Fig.~\ref{fig:loop-octet-decuplet} was presented in Ref.~\cite{Wang:2020hkn}.
In this section, we summarise those results, giving the splitting functions as a function of the light-cone variable $y=k^+/p^+$, where $k^\mu$ is the four-momentum of the kaon and $p^\mu$ is the four-momentum of the external proton. 
Following Refs.~\cite{Salamu:2018cny, Salamu:2019dok}, in our numerical calculation we choose for simplicity the Fourier transformation of the correlation function $F(a)$ in the nonlocal Lagrangian to take a dipole form,  
\begin{equation}
\widetilde{F}(k)
= \bigg(\frac{\Lambda^2-m_\phi^2}{\Lambda^2-k^2}\bigg)^2, 
\end{equation}
where $\Lambda = \Lambda_{B,T}$ are the cutoff parameters for the octet baryon-meson and decuplet baryon-meson vertices, and $m_\phi$ is the meson mass.

For the octet baryon rainbow diagram of Fig.~\ref{fig:loop-octet-decuplet}(a), the splitting function $\Delta f_{B\phi}^{(\rm rbw)}$ can be expressed as a sum of the on-shell, off-shell and $\delta$-function contributions,
\begin{equation}
\label{eq:Delta_BK_rbw}
\Delta f_{B\phi}^{(\rm rbw)}(y)
= \frac{C_{B\phi}^2 \overline{M}_{\!B}^2}{(4\pi f)^2}
  \left[ \Delta f_B^{\rm (on)}(y) + \Delta f_B^{(\rm off)}(y) 
  + \Delta f_B^{(\delta)}(y)
  \right],
\end{equation}
where $M$ is the nucleon mass and $\overline{M}_{\!B} = M_B+M$.
The coupling constants $C_{B\phi}$ are given in terms of the usual SU(3) coefficients $D$ and $F$.
The on-shell function is written as
\begin{eqnarray}\label{eq:fH-on}
\Delta f^{(\mathrm{on})}_B(y) 
&=& \overline{\Lambda}_B^8 
 \int\!\dd{k}^2_\perp\,
    \frac{y\, \big[\!-\!k^2_\perp + (\Delta_B + y M)^2 \big]
            \big( 4D_{B\phi} + D_{B\Lambda_B} \big)}
         {\yb^2 D^2_{B\phi}D^5_{B\Lambda_B}} \, ,
\end{eqnarray}
where $\yb=1-y$ is the light-cone fraction carried by the baryon, and we define for shorthand $\Delta_B = M_B-M$ and $\overline{\Lambda}_B^2 = \Lambda_B^2 - m_\phi^2$.
The functions $D_{B\phi}$ and $D_{B\Lambda_B}$ are defined as
\begin{subequations}
\begin{equation}
\label{eq:D_Bphi}
D_{B\phi}
= - \frac{1}{\yb} 
    \big( k^2_\bot + y M_B^2 + \yb\, m_\phi^2 - y \yb\, M^2 \big),
\end{equation}
and
\begin{equation}
\label{eq:D_BLambda}
\hspace*{-0.3cm}
D_{B\Lambda_B} 
= - \frac{1}{\yb}
    \big( k^2_\bot + y M_B^2 + \yb\, \Lambda_B^2 - y \yb\, M^2 \big).
\end{equation}
\end{subequations}
In the limit where $\Lambda_B \to \infty$, the local on-shell splitting function reduces to 
\begin{eqnarray}\label{eq:fH-on-local}
\Delta f^{(\mathrm{on})}_B(y) 
&=& \int\!\dd{k}^2_\perp\,
    \frac{y\, \big[\!-\!k^2_\perp + (\Delta_B + y M)^2 \big]}
         {\yb^2 D^2_{B\phi}},
\end{eqnarray}
which coincides with the results obtained in Refs.~\cite{Holtmann:1996be, Melnitchouk:1996fj, Melnitchouk:1999mv} in this limit.
The off-shell splitting function in Eq.~(\ref{eq:Delta_BK_rbw}) is given by
\begin{equation}
\label{eq:Deltaf_B-off}
\Delta f_B^{(\mathrm{off})}(y)
= \frac{2\overline{\Lambda}_B^8}{\overline{M}_{\!B}\!}
  \int\!\dd{k}_\perp^2\,
  \frac{\big( \Delta_B + y M \big)}
       {\yb\, D_{B\phi} D_{B\Lambda}^4}.
\end{equation}
For the $\delta$-function term, $\Delta f_\phi^{(\delta)}$, which arises from meson loops with zero light-cone momentum ($k^+ = 0$), one has
\begin{equation}
\Delta f_B^{(\delta)}(y)
= \frac{1}{6\overline{M}^2_{\!B}}
  \bigg( 2 \Lambda_B^2 
        + 3 m_\phi^2 
            \Big( 1 + 2 \log\frac{m_\phi^2}{\Lambda_B^2} \Big)
        - \frac{6 m_\phi^4}{\Lambda_B^2}
        + \frac{m_\phi^6}{\Lambda_B^4}
  \bigg)\,
\delta(y).
\label{eq:fKdel}
\end{equation}

The splitting functions of the tadpole diagram in Fig.~\ref{fig:loop-octet-decuplet}(b) for the charged and neutral kaon loop contributions are given by
\begin{equation}
\Delta f_{K^+}^{\rm (tad)}(y) = \Delta f_{K^0}^{\rm (tad)}(y)
\equiv - \frac{\overline{M}_{\!B}^2}{(4\pi f)^2}\,
    \Delta f_\phi^{(\delta)}(y) \, ,
\end{equation}
where the generic tadpole function $\Delta f_\phi^{(\delta)}$ is related to the $\delta$-function term in the rainbow diagram in Eq.~(\ref{eq:fKdel}),
\begin{equation}
\Delta f_\phi^{(\delta)}(y) = - \Delta f_B^{(\delta)}(y) \, .
\end{equation}
The splitting function of the Kroll-Ruderman diagrams in Fig.~\ref{fig:loop-octet-decuplet}(c) can be written in terms of the off-shell and $\delta$-function contributions as
\begin{eqnarray}\label{eq:KR}
\Delta f^{\rm (KR)}_{B\phi}(y)
&=& \frac{C_{B\phi}\, \overline{M}_{\!B}^2}{(4\pi f)^2}\,
 \Big[ \Delta f_B^{(\rm off)}(y) + 2 \Delta f_B^{(\delta)}(y) \Big],
\end{eqnarray}
with the off-shell function $\Delta f_B^{(\rm off)}$ given in Eq.~(\ref{eq:Deltaf_B-off}) and the $\delta$-function component, $\Delta f_B^{(\delta)}$, in Eq.~(\ref{eq:fKdel}). 
Note that Eq.~(\ref{eq:KR}) has the opposite sign relative to Eq.~(49) of Ref.~\cite{Wang:2020hkn} because the coefficient $C_{B\phi}$ used here has the opposite sign compared to that in \cite{Wang:2020hkn}.

For the decuplet intermediate states, because of the higher spin of the baryon the polarized splitting functions are somewhat more complicated.
As in the octet case, the splitting function associated with the decuplet rainbow diagram in Fig.~\ref{fig:loop-octet-decuplet}(d) can be decomposed as a sum of on-shell, off-shell and $\delta$-function contributions,
\begin{equation}\label{eq:f-TK-rbw}
\Delta f_{T\phi}^{(\rm rbw)}(y)
= \frac{C_{T\phi}^2 \overline{M}_{\!T}^2}{(4\pi f_\phi)^2}\,
\Big[ \Delta f^{(\rm on)}_T(y)
    + \Delta f^{(\rm off)}_T(y)
    + \Delta f_T^{(\delta)}(y)
\Big],
\end{equation}
where $\overline{M}_{T}= M_T+M$. 
The couplings $C_{T\phi}$ are given in terms of the coefficient $\cal C$.
In our analysis, we will take ${\cal C} = - 2 D$ from SU(6) symmetry.
The on-shell part of the splitting function is given by
\begin{eqnarray}\label{eq:fT-on}
\Delta f^{(\rm on)}_T(y)  
&=& -\frac{\overline{\Lambda}_T^8}{18 M_T^2 \overline{M}_{\!T}^2} 
    \int\!\dd{k}^2_\bot\,
    \Big[ k^2_\bot + \big(\overline{M}_{\!T} - y M \big)^2 
    \Big]
    \Big[ k^4_\bot - 8 \yb M M_T\, k^2_\bot - \big(M_T^2-\yb^2 M^2\big)^2
    \Big]
\notag  \\
& & \hspace*{3.5cm} \times\,
    \frac{y\, \big( 4D_{T\phi} + D_{T\Lambda_T} \big)}
         {\yb^4 D_{T\phi}^2 D_{T\Lambda_T}^5},
\end{eqnarray}
where $\overline{\Lambda}_T^2 = \Lambda_T^2 - m_\phi^2$, and $D_{T\phi}$, $D_{T\Lambda}$ are defined in analogy with Eq.~(\ref{eq:D_Bphi}),
\begin{subequations}
\begin{eqnarray}
\label{eq:D_Tphi}
D_{T\phi}& =& - \frac{k^2_\bot + y M_T^2 + \yb\, m_\phi^2 - y \yb\, M^2}{\yb}, \\
D_{T\Lambda_T}& =& - \frac{k^2_\bot + y M_T^2 + \yb\, \Lambda_T^2 - y \yb\, M^2}{\yb} \, .
\end{eqnarray}
\end{subequations}
The off-shell decuplet function part is given by
\begin{eqnarray}\label{eq:fT-off}
\Delta f^{(\rm off)}_T(y)
&=&\frac{\overline{\Lambda}_T^8}
        {\big(3 M_T^2 \overline{M}_{\!T} \big)^2} 
\int \dd{k}^2_\bot\,
\frac{1}{\yb^3 D_{T\phi} D^4_{T\Lambda_T}}
\Big[ k^6_\bot
    + \big(\yb^2 M^2 - M_T^2 - 3 \yb M_T M\big)\, k^4_\bot 
\nonumber\\
&& \hspace*{2cm}
    -\, \big( 3 M_T^4 + 2 \yb M_T^3 M + 4 \yb^2 M_T^2 M^2 
            + 6 \yb^3 M_T M^3 + \yb^4 M^4 
        \big)\, k^2_\bot 
\nonumber\\
&& \hspace*{2cm}
    -\, \big( M_T^3 - 2 \yb M_T^2 M + \yb^3 M^3 \big)
      \big( M_T + \yb M \big)^3
\Big].
\end{eqnarray}
For the $\delta$-function contribution, we perform the $k_\perp$ integration analytically to obtain
\begin{eqnarray}
\label{eq:fT-del}
\Delta f^{(\delta)}_T(y)
&=& \frac{1}{\big(3 M_T \overline{M}_{\!T}\big)^2}
\bigg\{
\frac{\overline{\Lambda}_T^2}{12 \Lambda_T^4 M_T^2}
\Big[
4 \Lambda_T^8
+ \Lambda_T^6\, \Big(29M_T \overline{M}_{\!T} + M_TM+14M^2-14m_\phi^2 \Big)
\nonumber\\
&& 
+\, \Lambda_T^4\,
    \Big( 22 m_\phi^4 - 5 m_\phi^2 \big(14 M^2+30 M M_T+29 M_T^2\big)
        +\, 4 \overline{M}_{\!T}^2 \big(2 M^2+2 M M_T-3 M_T^2\big)
    \Big)
\nonumber\\
&& 
-\, 2 \Lambda_T^2\, m_\phi^2 
    \Big( m_\phi^2 \big(14 M^2+30 M M_T+29 M_T^2\big)
        -\, 5 \big(2 M^2+2 M M_T-3 M_T^2\big) \overline{M}_{\!T}^2
    \Big)
\nonumber\\
&& 
-\, 2 m_\phi^4 \overline{M}_{\!T}^2 
    \Big( 2 M^2 + 2 M M_T - 3 M_T^2 \Big)
\Big]
\nonumber\\
&+& \frac{1}{2M_T^2}
\Big[ 2 \overline{M}_T^2 \big(2 M^2+2 M M_T-3 M_T^2\big)
    - m_\phi^2 \big(14 M^2+30 M M_T+29 M_T^2\big)
    + 2 m_\phi^4   
\Big]
\nonumber\\
& & \hspace*{1cm} \times\,
m_\phi^2 \log\frac{m_\phi^2}{\Lambda_T^2}
\bigg\}\
\delta(y)\, .
\end{eqnarray}

For the octet-decuplet rainbow transition diagrams in Fig.~\ref{fig:loop-octet-decuplet}(e), the splitting function can also be written as a sum of three terms,
\begin{equation}
\label{eq:f-TBK}
\Delta f_{TB\phi}^{\rm (rbw)}(y)
= -\frac{C_{T\phi} C_{B\phi} \overline{M}_{\!T} \overline{M}_{\!TB}}{(4\pi f)^2}
\Big[
  \Delta f^{(\rm on)}_{TB}(y)
+ \Delta f^{(\rm off)}_{TB}(y)
+ \Delta f_{TB}^{(\delta)}(y)
\Big],
\end{equation}
where $\overline{M}_{TB} = M_T + M_B$.
Explicit calculation of the on-shell octet-decuplet transition function in Eq.~(\ref{eq:f-TBK}) gives
\begin{eqnarray}
\label{eq:fTB-on}
\Delta f^{(\rm on)}_{TB}(y)
&=& \frac{\overline{\Lambda}_B^4\overline{\Lambda}_T^4}
         {3 M_T^2 \overline{M}_{\!TB} \Delta_{TB}}
\int \frac{\dd{k}_\bot^2}{\yb^2}\,          
\left(\frac{1}{D_{T\phi}D^2_{T\Lambda_B}D^2_{T\Lambda_T}}-\frac{1}{D_{B\phi}D^2_{B\Lambda_B}D^2_{B\Lambda_T}}\right)
       \nonumber\\
&& \times
\Big[
    k_\bot^4
    - \yb M (3M_T-M_B)\, k_\bot^2 
    - 2 M_T\Delta_{TB}\, k_\bot^2 
\nonumber\\ 
&& \quad
    -\, \big(\Delta_B+yM\big) \big(\Delta_T+yM\big) \big(\overline{M}_{\!T}-yM\big)^2
\Big],
\end{eqnarray}
where $\Delta_{TB} = M_T - M_B$, while for the off-shell transition function we have
\begin{eqnarray} 
\label{eq:fTB-off}
\Delta f^{(\rm off)}_{TB}(y)  
&=& \frac{\overline{\Lambda}_B^4\overline{\Lambda}_T^4}
         {3 M_T^2 \overline{M}_{\!T} \overline{M}_{\!TB}} 
\int\frac{\dd{k}_\bot^2}{\yb^2} 
\Bigg\{
\frac{\overline{M}_{\!T} 
      \big( k_\bot^2 (\yb M+2M_T) - (M_T-\yb M) (\overline{M}_{\!T}-yM)^2
      \big)}
     {D_{T\phi} D^2_{T\Lambda_B} D^2_{T\Lambda_T}}
\nonumber\\
&+&\frac{1}{D_{B\phi} D^2_{B\Lambda_B} D^2_{B\Lambda_T}}
\Big[
  k_\bot^4
+ \big( 
    \yb M (3M+4M_B) + 3(1\!+\!\yb) M_B M_T + M M_T\!-\!2 M_T^2
  \big) k_\bot^2 
\nonumber\\
&& \quad
-\, (M_B-\yb M) 
  \Big( 
    y^3 M^3 - y^2 M \big( 2 M_B M_T + (4M+M_B) \overline{M}_{\!T} \big)
\nonumber\\
&& \qquad\qquad\qquad\quad
    -\, y\, \big( M_B^2 (M+3M_T) 
                + M M_B \Delta_T 
                - M (5M-M_T) \overline{M}_{\!T}
          \big)
\nonumber\\
&& \qquad\qquad\qquad\quad
    +\, M_B^3 
        + M_T^3 
        + 2M \Delta_B \overline{M}_{\!B} 
        + M_T(2M+M_B) \overline{M}_{\!TB}\Big)
\Big]
\Bigg\}.
\end{eqnarray}
Finally, for the $\delta$-function contribution to the octet-decuplet transition, we have 
\begin{eqnarray}
\label{eq:fTB-delta}
\Delta f^{(\delta)}_{TB}(y)
&=& \frac{\delta(y)}
    {6 M_T^2 \overline{M}_T \overline{M}_{TB} \big(\Lambda_B^2-\Lambda_T^2\big)^3}
\Bigg\{
2 \big(\Lambda_B^2+\Lambda_T^2\big)
    \overline{\Lambda}_B^4  \overline{\Lambda}_T^4\,
    \log\frac{\Lambda_B^2}{\Lambda_T^2}
\nonumber\\
&+& \overline{\Lambda}_T^4 \log\frac{m_\phi^2}{\Lambda_B^2}\, 
\Big[
2 m_\phi^2\, 
\big( 
    (\Lambda_B^2+\Lambda_T^2)
    (\overline{M}_T \overline{M}_{TB} + \Delta_B \overline{M}_{B})
    - \Lambda_B^2 \Lambda_T^2
    - 2 \Lambda_B^4
\big)
\nonumber\\
&& \qquad\qquad\,
+\, 2 m_\phi^4\, (\Lambda _B^2+\Lambda_T^2)
- \Lambda_B^4
  \big( 4 \overline{M}_{TB} \overline{M}_T 
        + 4\Delta_B\overline{M}_B 
        - \Lambda_B^2
        - \Lambda_T^2
        \big)
\Big]
\nonumber\\
&-& \overline{\Lambda}_B^4 \log \frac{m_{\phi }^2}{\Lambda_T^2}\,
\Big[
2 m_\phi^2 
\big( 
    (\Lambda_B^2+\Lambda_T^2)
    (\overline{M}_T\overline{M}_{TB} + \Delta_B \overline{M}_{B})
    - \Lambda_B^2 \Lambda_T^2
    - 2 \Lambda_T^4
\big)
\nonumber\\
&& \qquad\qquad\,
+\, 2 m_\phi^4 
    (\Lambda_B^2+\Lambda_T^2)
-   \Lambda_T^4 
    \big( 4 \overline{M}_{TB} \overline{M}_T
        + 4 \overline{M}_B \Delta_B
        - \Lambda_B^2
        - \Lambda_T^2
    \big)
\Big]
\nonumber\\
&+& (\Lambda_B^2-\Lambda_T^2) \overline{\Lambda}_B^2 \overline{\Lambda}_T^2\,
\Big[
    m_\phi^2 
    \big( 4 \overline{M}_{TB}\overline{M}_T
        + 4\Delta_B\overline{M}_B
        + \Lambda_B^2
        + \Lambda_T^2
    \big) 
\nonumber\\
&& \qquad\qquad\qquad\quad
-\, 2 (\Lambda_T^2+\Lambda _B^2)
      (\overline{M}_{TB}\overline{M}_T + \Delta_B\overline{M}_{B})
- 2 \Lambda_B^2 \Lambda_T^2
\Big]
\Bigg\}\, .
\end{eqnarray}
In the local limit, where the regulator parameters $\Lambda_{B,T} \to \infty$, all of the splitting functions presented here are consistent with those obtained using Pauli-Villars (PV) regularization in Ref.~\cite{Wang:2020hkn}.

Using the above set of splitting functions, the strange quark PDF can be computed in the form of convolutions with the strange quark PDFs in the hadronic configurations in terms of the explicit hadronic configurations as~\cite{Wang:2020hkn, Salamu:2019dok}
\begin{eqnarray}
\label{eq:convolution-1}
\Delta s(x)
&=& \sum_{B\phi}
    \Big( \Delta \bar{f}_{B\phi}^{(\rm rbw)} \otimes \Delta s_B
        + \Delta \bar{f}_{B\phi}^{(\rm KR)}  \otimes \Delta s_B^{(\rm KR)}
    \Big)
 +\ \sum_\phi \Delta \bar{f}_\phi^{(\rm tad)} \otimes \Delta s_\phi^{(\rm tad)}
\nonumber\\
&+& \sum_{T \phi}
    \Delta \bar{f}_{T \phi}^{(\rm rbw)} \otimes \Delta s_{T}
    +\sum_{T B \phi} \Delta \bar{f}_{TB \phi} \otimes \Delta s_{T B}\ ,   
\end{eqnarray}
where the symbol ``$\otimes$'' represents a convolution integral, and for notational convenience we define the splitting functions
    $\Delta \bar{f}_j(y) \equiv \Delta f_j(\yb)$.
The input PDFs for the strange quark in the hadronic configurations on the right-hand side of Eq.~(\ref{eq:convolution-1}) can be related to the valence quark unpolarized $u(x)$ and $d(x)$ PDFs and polarized $\Delta u(x)$ and $\Delta d(x)$ PDFs in the proton by comparing the coefficients of the axial-vector operators using SU(6) symmetry~\cite{Wang:2020hkn}.
Specifically, the input PDFs $\Delta s_B$ in the rainbow diagram with intermediate octet states are related to the polarized PDFs in the proton as
\begin{eqnarray}
\Delta s_\Lambda(x)
&=& \frac13 \big( 2\Delta u(x)-\Delta d(x) \big),~~~~~
\Delta s_{\Sigma^+}(x)
 = \Delta s_{\Sigma^0}(x)
 = \Delta d(x).
\end{eqnarray}
The input PDFs in the Kroll-Ruderman diagrams, $\Delta s_B^{(\rm KR)}$, are related to the unpolarized distributions in the proton by
\begin{eqnarray}
\Delta s_\Lambda^{(\rm KR)}(x)
&=& \frac{1}{2\sqrt3} \big( 2u(x) - d(x) \big),~~~~~
\Delta s_{\Sigma^+}^{(\rm KR)}(x)
 = \sqrt{2}\Delta s_{\Sigma^0}^{(\rm KR)}(x)
 = \frac{1}{\sqrt2} d(x).
\end{eqnarray}
For $\Delta s_\phi^{(\rm tad)}$ in the tadpole diagram, we have the relationship
\begin{eqnarray}
\Delta s_{K^+}^{(\rm tad)}(x) = \frac12 \Delta u(x),~~~~~
\Delta s_{K^0}^{(\rm tad)}(x) = \frac12 \Delta d(x).
\end{eqnarray}
Finally, for the decuplet $\Delta s_T$ and octet-decuplet transition $\Delta s_{TB}$, these can be expressed as
\begin{eqnarray}
\Delta s_{\Sigma^{*+}}(x)
&=& \Delta s_{\Sigma^{*0}}(x)
 =  \frac12 \big( \Delta u(x) - 2 \Delta d(x) \big), \\
\Delta s_{\Sigma^{*+} \Sigma^+}(x)
&=& -\Delta s_{\Sigma^{*0} \Sigma^0}(x)
 =  \frac{1}{\sqrt3} \big( \Delta u(x) - 2 \Delta d(x) \big).
\end{eqnarray}
With these inputs, we can proceed to compute the strange helicity PDF in the proton in Eq.~(\ref{eq:convolution-1}) numerically, as we discuss next in the following section.

\clearpage
\section{Numerical results}
\label{sec:results}

In this section we discuss the numerical results for the spin-dependent strange quark distributions in the proton from the present analysis.
In the previous analysis of meson loop contributions to the spin-averaged strange quark PDFs in the proton~\cite{Salamu:2018cny, Salamu:2019dok}, the regulator parameter $\Lambda$ was determined by fitting the cross section data for inclusive baryon production in high-energy $pp$ scattering, $pp \to BX$, for different species of baryon $B$.
The best fit yielded the values $\Lambda_B = 1.1(1)$~GeV and $\Lambda_T = 0.8(1)$~GeV for the octet and decuplet intermediate cases, respectively. 
In the present analysis of spin-dependent PDFs we use the same parameters $\Lambda_B$ and $\Lambda_T$ and the coupling constants $C_{B\phi}$ and $C_{T\phi}$ to compute the splitting functions numerically.

\subsection{Splitting functions}

\begin{figure}[t]
\begin{minipage}[b]{.45\linewidth}
\hspace*{-0.55cm}\includegraphics[width=1.125\textwidth, height=5.5cm]{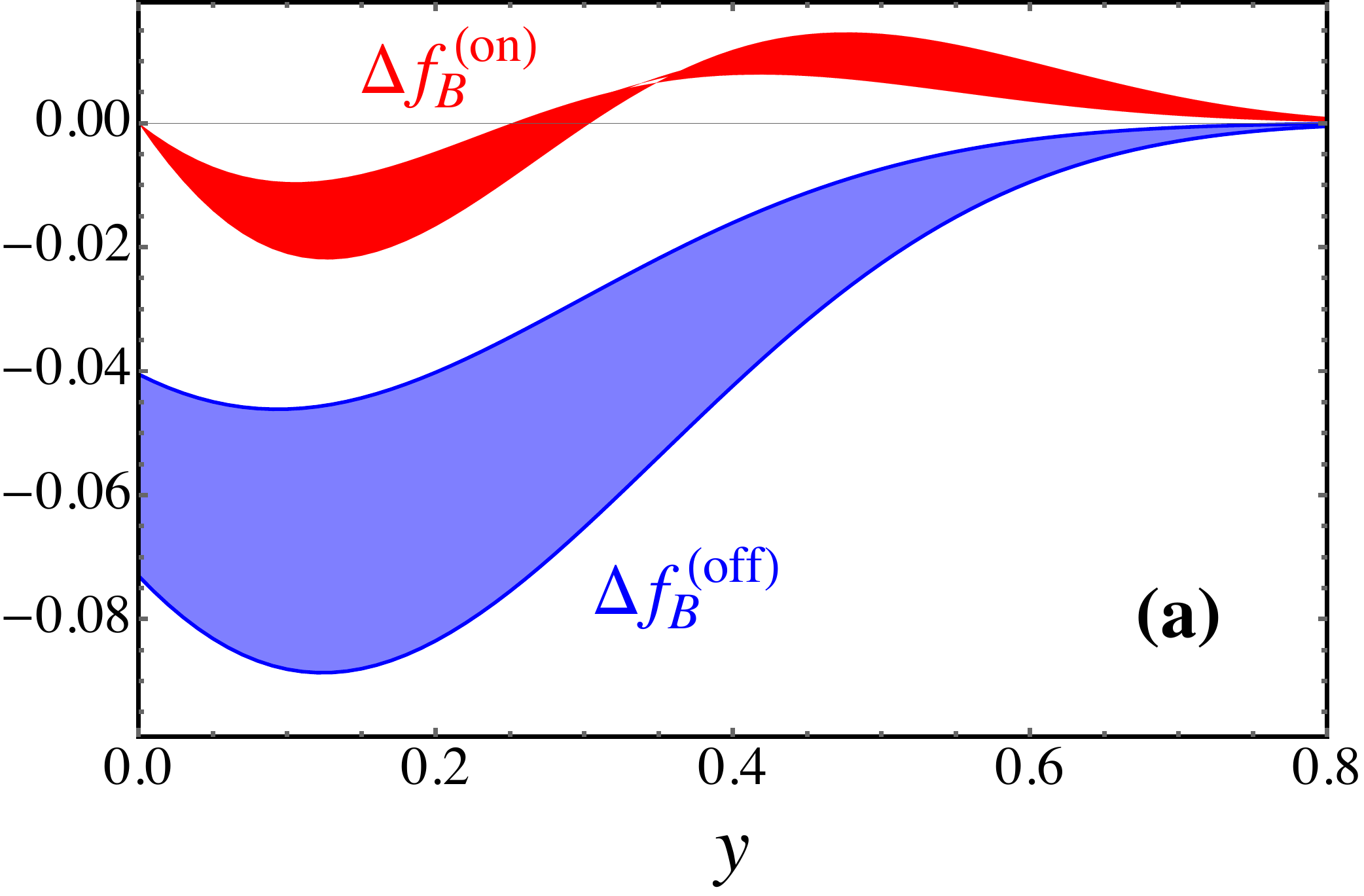} 
 \vspace{-22pt}
\end{minipage}
\hfill
\begin{minipage}[b]{.45\linewidth}   
\hspace*{-1.1cm} \includegraphics[width=1.1\textwidth, height=5.5cm]{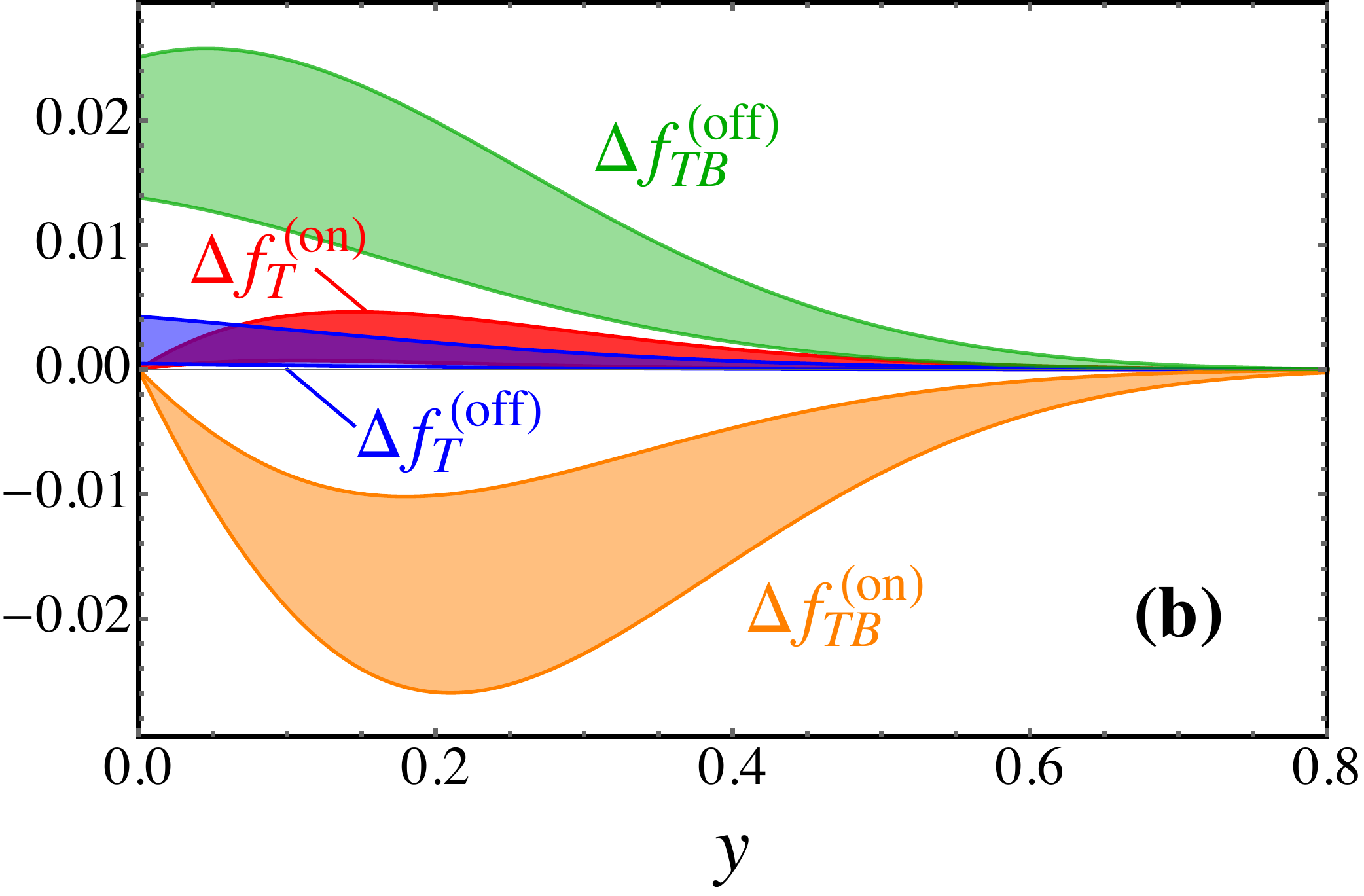}
   \vspace{0pt}
\end{minipage} 
\caption{On-shell ``(on)'' and off-shell ``(off)'' contributions to the spin-dependent splitting functions in the nonlocal EFT for {\bf (a)} the octet baryon $\Delta f_B$ and {\bf (b)} the decuplet baryon $\Delta f_T$ and octet-decuplet interference $\Delta f_{TB}$ intermediate states.  The bands correspond to regulator parameter values $\Lambda_B = 1.0 - 1.2$~GeV for octet and $\Lambda_T = 0.7 - 0.9$~GeV for decuplet baryons.}
\label{f:splitting-functions}
\end{figure}

The on-shell and off-shell contributions to the spin-dependent splitting functions in the nonlocal EFT calculation for the strange octet, decuplet and octet-decuplet baryon interference intermediate states are shown in Fig.~\ref{f:splitting-functions}, with the bands corresponding to regulator cutoff values $\Lambda_B = 1.0 - 1.2$~GeV and $\Lambda_T = 0.7 - 0.9$~GeV for the octet and decuplet baryons, respectively.
The results are qualitatively similar to those found for the splitting functions in the local EFT calculation with Pauli-Villars regularization~\cite{Wang:2020hkn}, although the magnitude there was somewhat smaller.
For the octet baryon case, the on-shell splitting function $\Delta f_B^{(\rm on)}$ is negative at small meson momentum fractions $y$, but changes sign to become positive for $y \gtrsim 0.3$.
The off-shell contribution $\Delta f_B^{(\rm off)}$ remains negative for all $y$ values, and is $\approx 4$ times larger in magnitude at the peak $y \approx 0.1$ than the on-shell contribution.
Note that because of the mass difference between the hyperon and nucleon, the off-shell function is nonzero at $y=0$.
Interestingly, compared with the corresponding spin-averaged splitting functions from Ref.~\cite{Salamu:2018cny}, the spin-dependent off-shell function is identical, while the spin-averaged on-shell function is positive.

For intermediate state involving decuplet baryons, the splitting functions are generally smaller in magnitude than for the octet baryons.
Both the decuplet on-shell and off-shell splitting functions are positive, and as in the octet case the on-shell contributions vanish at $y=0$, while the off-shell contributions remain nonzero.
The octet-decuplet transition splitting functions are significantly larger in magnitude than the decuplet functions, indicating a greater role played by the $BT$ transition in spin-dependent observables than for the diagonal $T$ contributions.
The on-shell transition function is negative, peaking at $y \approx 0.2$, while the off-shell transition function is positive and decreases with $y$.
Compared with the corresponding splitting functions for the local EFT calculation with Pauli-Villars regularization~\cite{Wang:2020hkn}, the shapes are quite similar; however, the magnitude of the transition contributions in the nonlocal case are again larger.

\subsection{Strange quark polarization}

Using the convolution formula (\ref{eq:convolution-1}), the polarized strange quark PDF is evaluated in terms of the derived hadronic splitting functions and the PDFs in the the hadronic intermediate state configurations. 
The $\delta(y)$ terms in the splitting functions, although not shown in Fig.~\ref{f:splitting-functions}, make significant contributions to the strange quark PDFs via the convolutions.
With the SU(3) relations, the strange quark PDFs for the intermediate states can be given in terms of the spin-dependent and spin-averaged $u$ and $d$ quark PDFs in the proton \cite{Wang:2020hkn}, which are determined from global QCD analyses of high-energy polarized~\cite{Ethier:2017zbq, Nocera:2014gqa, Hartland:2012ia} and unpolarized~\cite{Ball:2017nwa, Sato:2019yez, MARATHON:2021vqu, Cocuzza:2021rfn} scattering data.
In this analysis, for the spin-averaged $u$ and $d$ quark distributions in the proton we use the CJ15 parametrization from Ref.~\cite{Accardi:2016qay}, while the polarized $\Delta u$ and $\Delta d$ PDFs are taken from the JAM analysis in Ref.~\cite{Ethier:2017zbq}.

\begin{figure}[t] 
\begin{minipage}[b]{.45\linewidth}
\hspace*{-0.3cm}\includegraphics[width=1.1\textwidth, height=5.5cm]{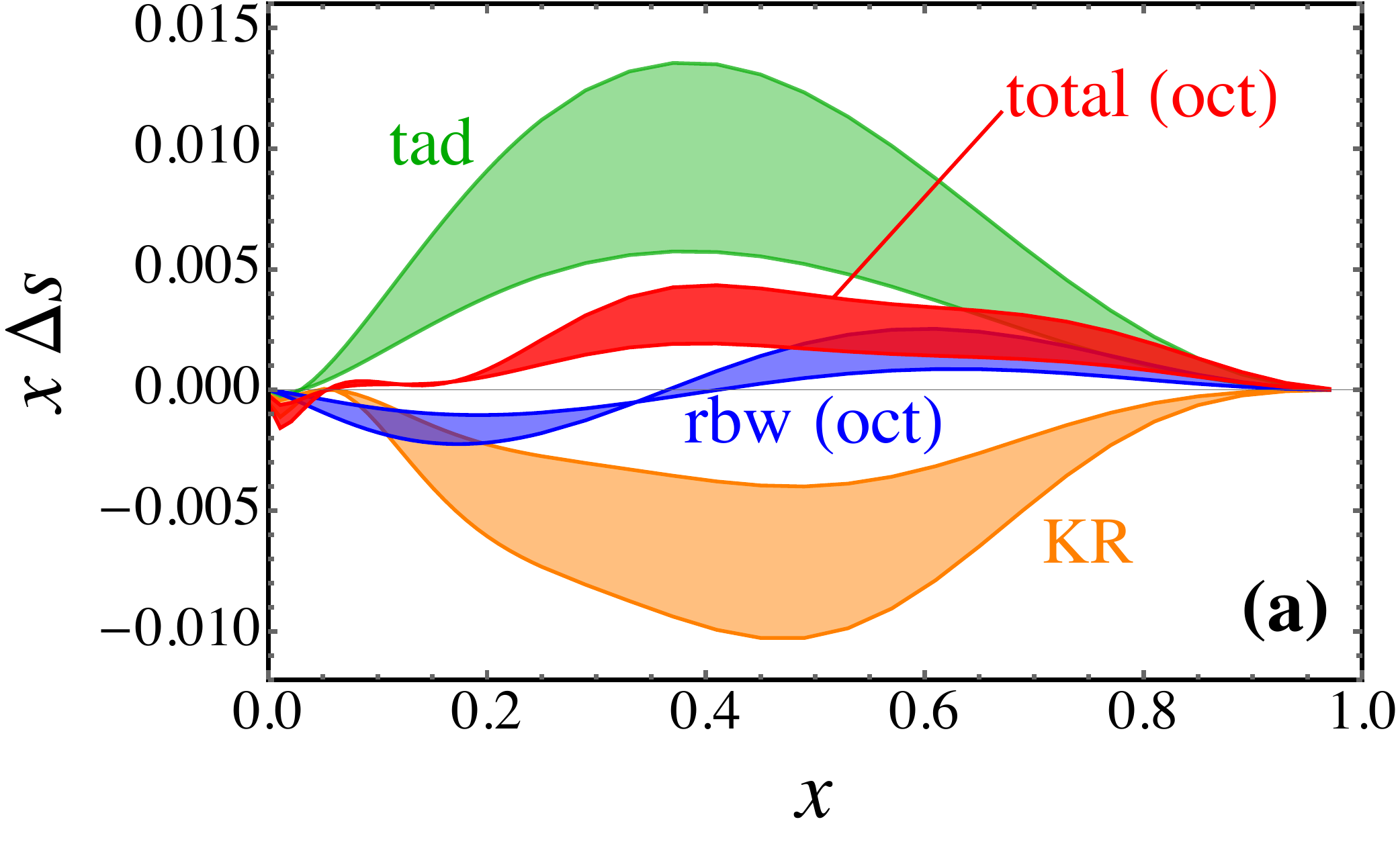}
\end{minipage}
\hfill
\begin{minipage}[b]{.45\linewidth}   
\hspace*{-0.95cm} \includegraphics[width=1.1\textwidth, height=5.5cm]{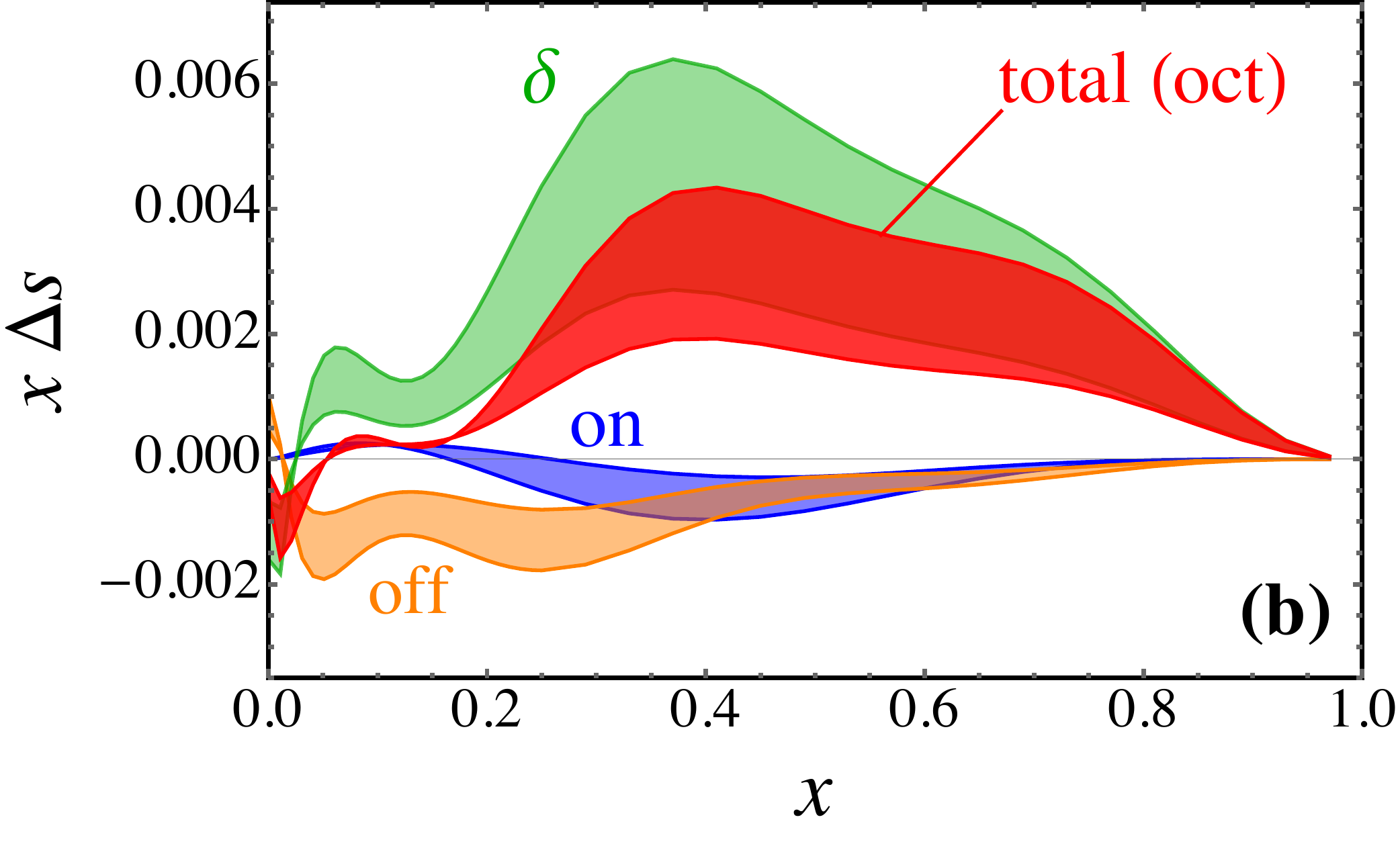} 
\end{minipage}  
\\[0.3cm]
\begin{minipage}[t]{.45\linewidth}
\hspace*{-0.5cm}\includegraphics[width=1.12\textwidth, height=5.5cm]{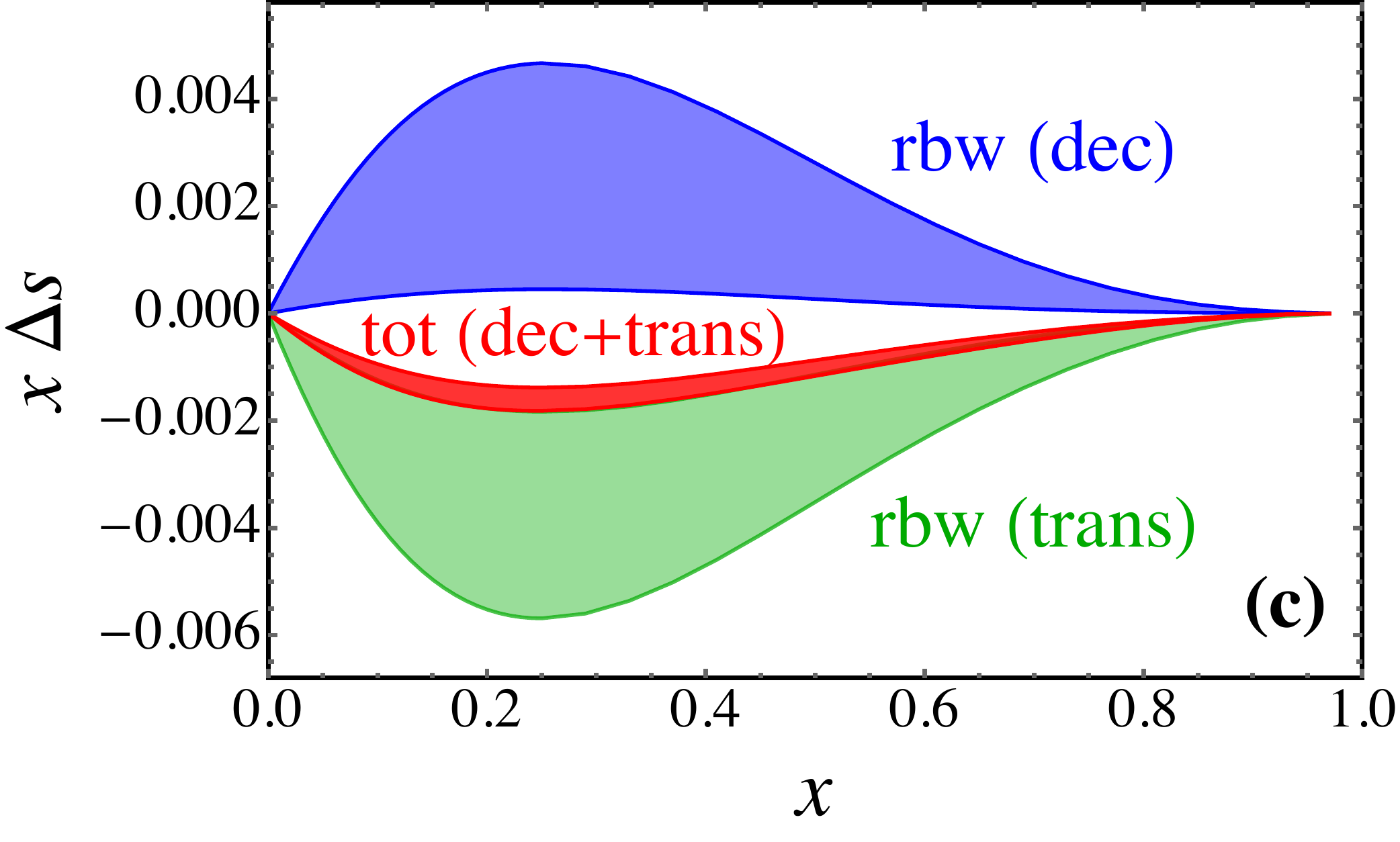}
\end{minipage}
\hfill
\begin{minipage}[t]{.45\linewidth}   
\hspace*{-0.85cm} \includegraphics[width=1.1\textwidth, height=5.5cm]{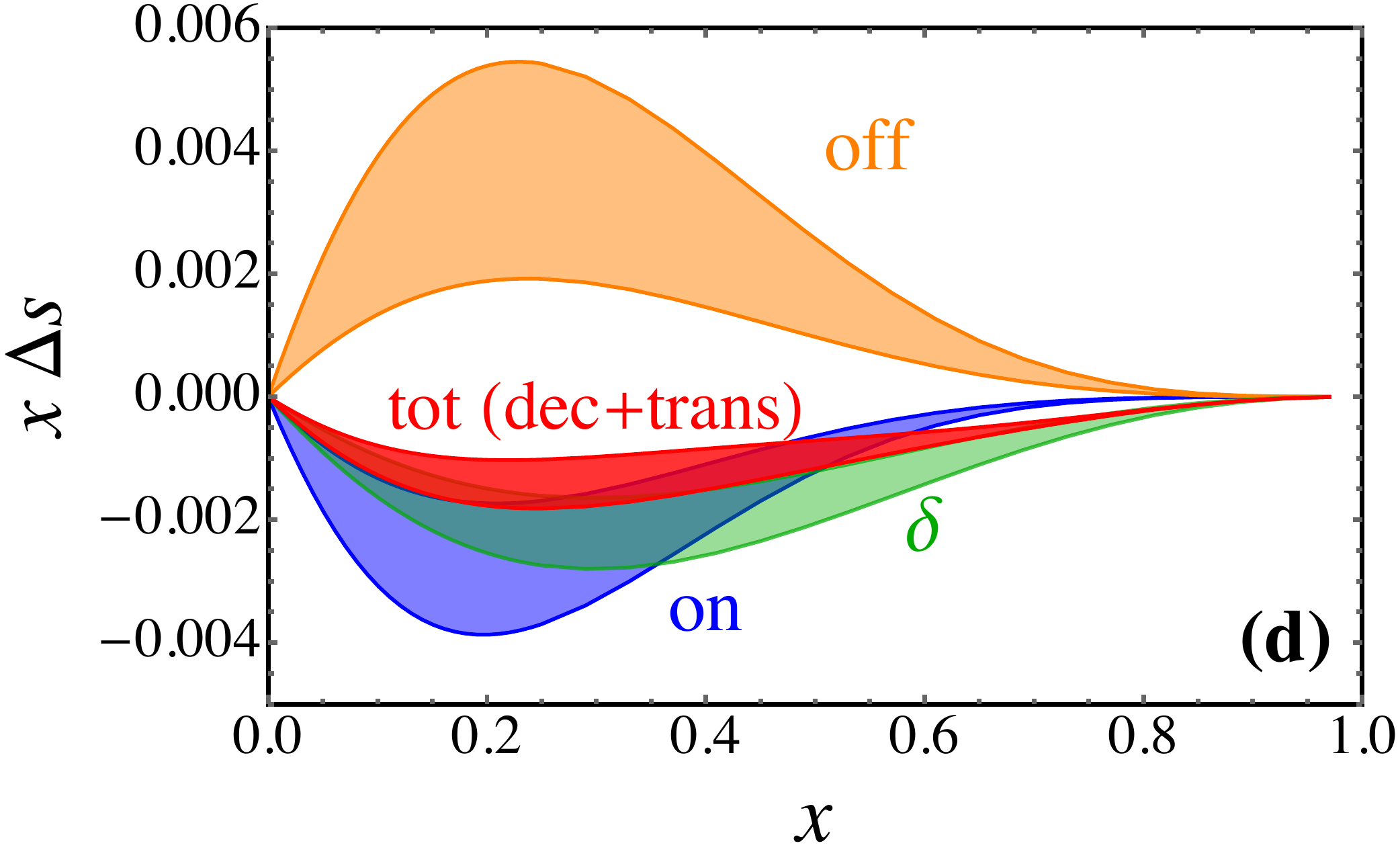}  
   \vspace{-12pt}
\end{minipage} 
\caption{Contributions to the strange quark polarization in the proton, $x \Delta s(x)$, at $Q^2=1$~GeV$^2$ from the octet rainbow, Kroll-Ruderman, and tadpole diagrams [{\bf (a), (b)}]; and the decuplet rainbow and decuplet-octet transition rainbow diagrams [{\bf (c), (d)}]. The uncertainty bands correspond to the range of $\Lambda_B = 1.0$--1.2~GeV for octet and $\Lambda_T = 0.7$--0.9~GeV for the decuplet and decuplet-octet transition. The left column [{\bf (a), (c)}] shows the decomposition according to the type of diagram, while the right column [{\bf (b), (d)}] shows the decomposition according to the type of splitting function.}
\label{fig:xs_decomp}
\end{figure}


The contributions to the polarized strange PDF $x\Delta s$ from the various terms in Eq.~(\ref{eq:convolution-1}) are shown in Fig.~\ref{fig:xs_decomp}, illustrating both decompositions in terms of diagram types and in terms of splitting function types.
In Fig.~\ref{fig:xs_decomp}(a) we observe large cancellations between contributions from the positive tadpole and negative KR diagrams.
In contrast, the magnitude of the octet rainbow diagram contribution is relatively small, changing sign from negative at small $x$ to positive at larger $x$.
The sum of these contribution is positive for $x \gtrsim 0.2$ with magnitude $\lesssim 0.004$.
Compared with the calculation in Ref.~\cite{Wang:2020hkn} which used PV regularization, the tadpole contribution in the nonlocal case is significantly larger, and the contribution from the KR diagram has the opposite sign. 
This is mainly because the $\delta$-function contribution in the nonlocal calculation is much larger, as can be seen in the decomposition in Fig.~\ref{fig:xs_decomp}(b).
Here, the (positive) $\delta$-function term gives the largest contribution, while the (mostly negative) on-shell and off-shell terms are somewhat smaller.
The on-shell contribution changes smoothly from positive at small $x$ to negative at $x \gtrsim 0.2$, whereas the off-shell term has largest magnitude at low $x$.
The overall behavior of the octet contribution is driven by the $\delta$-function contribution.

The contributions from diagrams involving decuplet baryons in the intermediate states are shown in Fig.~\ref{fig:xs_decomp}(c) and (d).
There are strong cancellations seen between the positive decuplet rainbow and the negative octet-decuplet transition contributions, resulting in a total result that is negative with a small magnitude ($\lesssim 0.002$), comparable to the total octet contribution.
Furthermore, in contrast to the octet case, the off-shell decuplet contributions are positive, but cancelled somewhat by the negative on-shell and $\delta$-function terms.
Compared with the earlier calculation~\cite{Wang:2020hkn} with PV regularization, the signs of each of the decuplet rainbow and octet-decuplet transition terms is the same, but somewhat different magnitudes result in a net contribution which is negative.

\begin{figure}[t]
\includegraphics[width=0.8\columnwidth]{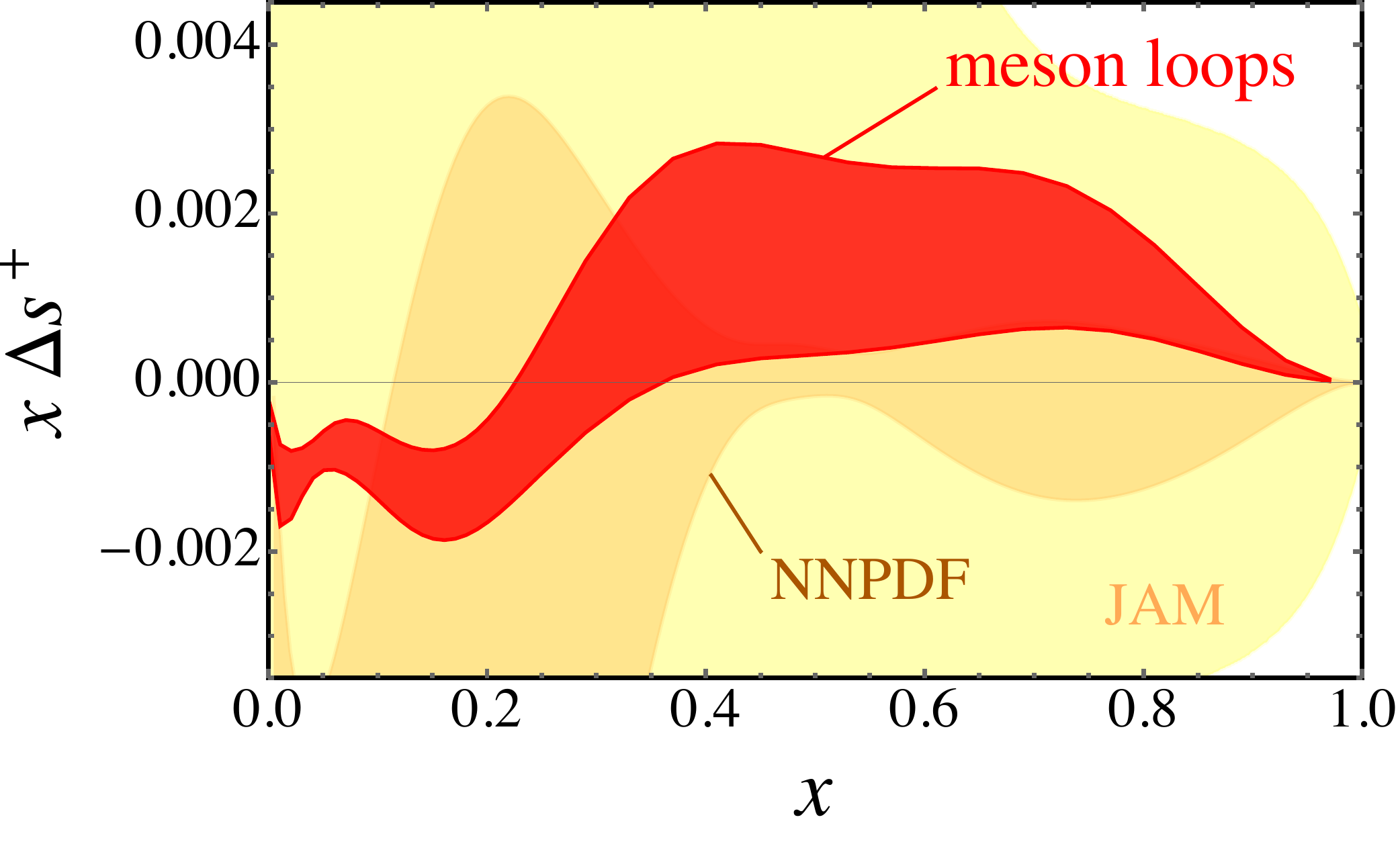}
\caption{Meson loop contribution to the polarized strange quark PDF (red band) 
    $x \Delta s^+ \equiv x \Delta s + x \Delta \bar s$ 
compared with the results from the NNPDF~\cite{Nocera:2014gqa, Hartland:2012ia} (orange band) and JAM~\cite{Ethier:2017zbq} (yellow band) global QCD analyses, at a scale $Q^2 = 1$~GeV$^2$. The meson loop band reflects the range of cutoff parameters $\Lambda_B=1.0$--1.2~GeV and $\Lambda_T=0.7$--0.9~GeV for the octet and decuplet sectors, respectively. }
\label{f.PDFs}
\end{figure}

In Fig.~\ref{f.PDFs} we plot the full results for the strange quark PDF, $x \Delta s^+(x)$, including all octet, decuplet and transition contributions (the contribution from meson loops to polarized antistrangeness $\Delta \bar s$ is zero). For comparison, the results for $x \Delta s^+(x)$ from the NNPDF~\cite{Nocera:2014gqa} and JAM~\cite{Ethier:2017zbq} global QCD analyses at $Q^2 = 1$~GeV$^2$ are also shown. The overall magnitude of the calculated strange polarization in the nonlocal chiral effective theory is relatively small, with $x \Delta s$ starting out negative at $x \lesssim 0.25$ and becoming positive at larger $x$ values. The large uncertainty on the PDF parametrizations reflect the weak constraints that currently exist on $\Delta s$ from data.
Other analyses in which the strange quark helicity-dependent distributions have been extracted have been under the assumption of $\Delta s = \Delta \bar{s}$ \cite{deFlorian:2008mr, Leader:2010rb, Borsa:2020lsz, Blumlein:2010rn, Khorramian:2010qa}. 
Similar to our result, the sign change of $x\Delta s$ from negative to positive with increasing $x$ was found in Refs.~\cite{deFlorian:2008mr, Leader:2010rb, Borsa:2020lsz}, although the zero appeared in the smaller-$x$ region. 
In some other extractions~\cite{Blumlein:2010rn, Khorramian:2010qa}, $\Delta s$ was negative in the entire $x$ region without any sign change, driven by the assumption of SU(3) flavor symmetry for the axial vector charges.
Recent lattice simulations of $x\Delta(s)$~\cite{Alexandrou:2020uyt} also suggest a sign change at $x \sim 0.3$, consistent with our result.

\begin{table}[tb]
\begin{center}
\caption{\label{tab:moment-octet}
Contributions from various diagrams to the integral of $\Delta s(x)$ at $Q^2 = 1$~GeV$^2$, in units of $10^{-2}$. The sum of the total octet, tadpole, and total decuplet and octet-decuplet transition terms is in the range
    \bm{$\langle \Delta s \rangle = [-0.51, -0.26] \times 10^{-2}$}.}
\begin{tabular}{c|ccccccc}     \hline\hline
~$\Lambda_B$ (GeV)                            &
~$\langle \Delta s \rangle_{B\, \rm rbw}^{(\rm on)}$ &
~$\langle \Delta s \rangle_{B\, \rm rbw}^{(\rm off)}$&
~$\langle \Delta s \rangle_{B\, \rm rbw}^{(\delta)}$ &
~$\langle \Delta s \rangle_{\rm KR}^{(\rm off)}$     &
~$\langle \Delta s \rangle_{\rm KR}^{(\delta)}$      &
~total octet~&
~$\langle \Delta s \rangle_{\rm tad}^{(\delta)}$~ 
\\ 
1.0     &
~~0.02  &
$-0.90$ &
~~0.73  &
~~0.87  &
$-1.55$ &
\bm{$-0.83$}&   
~~{\bf 0.84}
\\
1.1     &
~~0.00  &
$-1.38$ &
~~1.18  &
~~1.32  &
$-2.48$ &
\bm{$-1.36$}&   
~~{\bf 1.35}
\\
1.2     &
$-0.04$ &
$-1.94$ &
~~1.73  &
~~1.86  &
$-3.65$ &
\bm{$-2.04$}&   
~~{\bf 1.99}    
\\ \hline
~$\Lambda_T$ (GeV)                                   &
~$\langle \Delta s \rangle_{T\, \rm rbw}^{(\rm on)}$   &
~$\langle \Delta s \rangle_{T\, \rm rbw}^{(\rm off)}$  &
~$\langle \Delta s \rangle_{T\, \rm rbw}^{(\delta)}$   &
~$\langle \Delta s \rangle_{TB\, \rm rbw}^{(\rm on)}$  &
~$\langle \Delta s \rangle_{TB\, \rm rbw}^{(\rm off)}$  &
~$\langle \Delta s \rangle_{TB\, \rm rbw}^{(\delta)}$~ &
~~total decuplet~ 
\\  
0.7     &
~~0.04  &
~~0.02  &
~~0.04  &
$-0.45$ &
~~0.42  &
$-0.41$ &   
~\bm{$-0.34$}~
\\
0.8     &
~~0.14  &
~~0.08  &
~~0.18  &
$-0.83$ &
~~0.74  &
$-0.77$ &   
~\bm{$-0.46$}~ 
\\
0.9     &
~~0.33  &
~~0.22  &
~~0.53  &
$-1.24$ &
~~1.05  &
$-1.16$ &   
~\bm{$-0.27$}~  
\\ \hline\hline
\end{tabular}
\end{center}
\end{table}

Integrating the strange polarized PDFs over all $x$, the resulting contributions to the total moment $\langle \Delta s \rangle \equiv \int_0^1 \dd{x} \Delta s(x)$ are listed individually in Table~\ref{tab:moment-octet}.
This moment is especially interesting in view of its role in neutrino transport in neutron stars, which is a key cooling mechanism~\cite{Bollig:2017lki, Hobbs:2016xlg}.
With the exception of the on-shell component, the contributions from the octet states are larger than the corresponding ones from the decuplet baryons, with the total octet contribution several times larger than the total decuplet. 
For the range of $\Lambda_{B,T}$ values considered in this analysis, the octet intermediate state contribution to $\langle \Delta s \rangle$ is between $-0.0204$ and $-0.0083$, while the contribution from decuplet intermediate states is between $-0.0046$ and $-0.0027$.
The large positive contribution from the tadpole diagram cancels much of the (negative) octet and decuplet contributions, leaving a net strange quark polarization in the proton to be in the range
    $\langle \Delta s \rangle \approx [-0.0051, -0.0026]$.
Note that although the first moment of $\Delta s(x)$ is negative, the sign of the distribution is $x$ dependent. 
The overall negative value of $\langle \Delta s \rangle$ results from the relatively larger negative $\Delta s(x)$ at small $x$, including at $x=0$, compared with the smaller positive $\Delta s(x)$ at large $x$.

Interestingly, our results are comparable with those from Ref.~\cite{Wang:2020hkn} using PV regularization, where the lowest moment $\langle \Delta s \rangle$ was also negative and in the range $[-0.0050,-0.0025]$.
However, we should note that although the total moments $\langle \Delta s \rangle$ turn out to be similar, the individual contributions from the various terms are quite different.
In particular, the $\delta$-function terms for the intermediate octet states in the nonlocal case are significantly larger than those in the PV case. 
The calculated moment can also be compared with determinations from the JAM global QCD analysis~\cite{Ethier:2017zbq}, which yielded
    $\langle \Delta s^+ \rangle_{\scriptsize{\rm JAM}} = - 0.03(10)$.
Earlier phenomenological analyses generally found more negative values of the first moment of the strange quark polarization \cite{Leader:2010rb, Nocera:2014gqa, Khorramian:2010qa}, which were, however, largely driven by the assumption of SU(3) flavor symmetry for the axial vector charges.
%
Future data on semi-inclusive DIS and parity-violating inclusive DIS at the planned Electron-Ion Collider~\cite{Accardi:2012qut} should reduce the uncertainty on the extracted $\langle \Delta s^+ \rangle$, and allow a better discrimination between the $\Delta s$ and $\Delta \bar s$ distributions.

\clearpage
\section{Conclusion}
\label{sec:conclusion}

Our main aim in this paper was to examine the generation of polarized strangeness in the proton from meson loops computed within a nonlocal chiral effective field theory at the one loop level.
In contrast to previous calculations using local versions of the effective theory, the nonlocal implementation allows one to study hadron structure while {\it a priori} taking into account the finite size of the hadrons in a natural and consistent way.

We derived explicit expressions for the spin-dependent proton to pseudoscalar meson plus octet or decuplet baryon splitting functions, using a simple dipole shape for the Fourier transform of the correlation function that describes the hadrons' extended structure.
With the regulator parameters determined phenomenologically from spin-averaged measurements of semi-inclusive hyperon production in $pp$ collisions, the strange helicity PDF was computed from convolutions of the splitting functions and the parton distributions associated with the hadronic intermediate states.
%
The contributions involving octet baryons in the intermediate states were found to be several times larger in magnitude than those involving decuplet baryons, and significant cancellation was found between these and the tadpole contributions.
The result was a relatively small net polarized strange quark helicity, which was negative at small $x$ and positive at high $x$.

Integrated over $x$, the lowest moment was found to lie in the range
    $\langle \Delta s \rangle = [-5.1, -2.6] \times 10^{-3}$. 
Interestingly, this is very similar to the results found in the previous calculations~\cite{Wang:2020hkn} using a local effective chiral theory with Pauli-Villars regularization, even though the shape of the $\Delta s(x)$ distribution there was somewhat different.
This suggests that while the details of the meson loop calculation depend on the prescription chosen to regularize the short-distance behavior, the overall effect on the generated strange quark polarization is relatively robust. 

Our results are also qualitatively similar to those found in the recent global QCD analysis by the JAM collaboration,
    $\langle \Delta s^+ \rangle_{\scriptsize{\rm JAM}} = -0.03(10)$
\cite{Ethier:2017zbq}, as well as with the latest lattice QCD simulations from the ETM Collaboration,
    $\langle \Delta s^+ \rangle_{\scriptsize{\rm latt}} = -0.046(8)$
\cite{Alexandrou:2020sml}.
We expect uncertainties in the determination of $\langle \Delta s^+ \rangle$ on both of these fronts to decrease as more experimental and lattice data become available over the next few years.

\section*{Acknowledgments}

This work is supported by the Australian Research Council through the ARC Discovery Project DP180100497 (AWT), the DOE Contract No.~DE-AC05-06OR23177, under which Jefferson Science Associates, LLC operates Jefferson Lab, DOE Contract No.~DE-FG02-03ER41260, and by the NSFC under Grant No.~11975241.

\bibliographystyle{apsrev4-2}
\bibliography{ref_deltas}

\end{document}